# Unveiling the Shortwave Absorption Spectra of Alumina Aerosols: Implications for Solar Radiation Modification

Taveen Singh Kapoor[a,1], Prabhav Upadhyay[a,1], Jian Huang[b,1], Guodong Ren[b], John Cavin[b], Dhruv, Mitroo[c], Sandro Vattioni[d,e], John Dykema[f], Jan Sedlacek[e], Joshin Kumar[a], Jordan A. Hachtel[g], Lu Xu[a], Rohan Mishra[b,*], Rajan K. Chakrabarty[a,*]

[a] Center for Aerosol Science and Engineering, Department of Energy, Environmental and Chemical Engineering, Washington University in St. Louis, St. Louis, USA.

[b] Institute of Materials Science & Engineering, Department of Mechanical Engineering and Materials Science, Washington University in St. Louis, St. Louis, USA.

[c] Veterans Affairs Medical Center, St. Louis, USA.

[d] Institute for Atmospheric and Climate Science, ETH Zurich, Zurich, Switzerland

[e] Physikalisch-Meteorologisches Observatorium Davos and World Radiation Center, Davos, Switzerland.

[f] John A Paulson School of Engineering and Applied Sciences, Harvard University, Cambridge, MA, USA

[g] Center for Nanophase Materials Sciences, Oak Ridge National Laboratory, Oak Ridge, USA.

[1] These authors contributed equally

* Correspondence to Rajan K. Chakrabarty or Rohan Mishra

**Email:** chakrabarty@wustl.edu; rmishra@wustl.edu

**Abstract**

Alumina is proposed for Stratospheric Aerosol Injection (SAI)-based solar radiation modification due to its presumed ability to scatter sunlight strongly while absorbing weakly. Alumina is assigned negligible solar shortwave absorption in climate models; this assumption is not validated owing to technological challenges in quantifying its weak absorption signals. We report alumina's shortwave imaginary refractive index (*k*), a determinant of its absorption strength, using sensitive *in situ* photoacoustic spectrometry, finding values ranging from 1.4 x $10^{-4}$ to 1.2 x $10^{-3}$. Particle-scale electron energy-loss spectroscopy provided independent validation and revealed that the non-ideal absorption arises from oxygen vacancy defects in the alumina's crystal structure. Aerosol chemistry climate model simulations to evaluate shortwave absorption radiative effects revealed insignificant impacts on radiative forcing and stratospheric warming. Our findings indicate that alumina's shortwave absorption, previously reported as a source of uncertainty, is unlikely to affect SAI impact calculations.

**Plain Language Summary**

Solar radiation modification (SRM) based on Stratospheric Aerosol Injection (SAI) is proposed to reduce the impacts of global warming by reflecting sunlight. Recent studies have suggested solid particles such as alumina as candidates for SAI because they are expected to be strong reflectors and very weak absorbers of sunlight. However, measuring their weak sunlight absorption has been challenging, leaving a critical uncertainty in climate models. In this study, we used photoacoustic spectrometry to measure the absorption strength of alumina nanoparticles. We found that their absorption is non-negligible, being over a thousand times higher than previous theoretical estimates used in climate models. Electron microscopy studies coupled with previously reported first-principles density functional theory calculations reveal that this deviation from expected ideal behavior is likely caused by oxygen vacancies within alumina. These findings emphasize the need to revisit the optical properties of other solid particle materials proposed for SAI. In this study, when we incorporated the measured absorption into climate models, we found that solar absorption had little influence on SAI efficacy. Our study constrains this uncertainty, increasing confidence in alumina as an SAI candidate. However, other uncertainties surrounding alumina remain, particularly concerning its chemical reactivity and potential to deplete stratospheric ozone.

**Key Points**

- Measurements of alumina's presumed negligible shortwave absorption reveal finite imaginary refractive indices ($1.4 \times 10^{-4}$ to $1.2 \times 10^{-3}$).
- Absorption arises from imperfections in ideal crystal structure, indicating the need to investigate all solid particle candidates carefully.
- Constrained a source of stratospheric aerosol injection uncertainty: alumina's shortwave absorption is not a concern for radiative impacts.

## 1. Introduction

Stratospheric Aerosol Injection (SAI) is a form of Solar Radiation Modification (SRM, or Sunlight Reflecting Method) intended to mitigate the impacts of anthropogenically driven climate change, supplementing efforts to reduce greenhouse gas emissions (Budyko & Miller, 1974; Crutzen, 2006). This technique proposes the deliberate injection of reflective aerosol particles into the stratosphere, where their long residence times allow for sustained reflection of incoming solar radiation. The concept is inspired by the temporal tropospheric cooling effect of large explosive volcanic eruptions, which emit $SO_2$ into the stratosphere. After oxidation of $SO_2$, liquid particles with sulfuric acid and water are formed, which scatter solar radiation (Robock & Mao, 1995; McCormick et al., 1995; Parker et al., 1996). However, the liquid sulfate aerosols in the stratosphere can introduce significant drawbacks, including undesirable stratospheric warming due to longwave absorption, which can lead to perturbations of atmospheric circulation, the hydrological cycle as well as stratospheric ozone (Tilmes et al., 2008, Wunderlin et al., 2024). Consequently, research efforts are now focused on exploring alternative materials with inherently weaker absorption across both the infrared and shortwave spectra to minimize these side effects (Blackstock et al., 2009; Teller et al., 1996; Pope et al., 2012; Dykema et al., 2016). Among the several candidates proposed, alumina ($Al_2O_3$) has emerged as a particularly promising alternative (Dykema et al., 2016; IPCC, 2023; Weisenstein et al., 2015; Vattioni, Käslin, et al., 2024, Stefanetti et al., 2024). However, this view rests on an incomplete understanding of alumina's absorption properties in the shortwave solar spectrum (wavelength range: 300-3000 nm).

Alumina is a wide bandgap insulator (8.7 eV for the α-phase, ~142 nm). Below this bandgap, it is expected to exhibit negligible absorption, as described by the Urbach energy model (Moosmüller, H. 2011, Urbach, F. 1953). Photons with lower energy (i.e., longer wavelengths) cannot excite electrons across the bandgap, so an ideal crystal is not expected to absorb light. However, crystal defects—present in finite concentrations due to entropy—can introduce electronic states within the bandgap (Freysoldt et al., 2014). These localized states may enable sub-bandgap absorption by promoting electronic transitions from the valence to defect levels or from the defect levels to the conduction band, contributing to absorption in the shortwave region (Pustovarov et al., 2010) (Fig. 1). Despite this possibility, alumina's shortwave absorption has not been accurately measured due to technological limitations (Tomiki et al., 1993). In a widely used optics handbook, Tropf and Thomas (1997) compiled the imaginary refractive index (*k*) of *α*-alumina, the thermodynamically stable polymorph. *k* is an intrinsic property of a material governing its light absorption. While the handbook reports reliable data in the ultraviolet and infrared, shortwave values are conspicuously absent, leaving a zone of "no *k* data" from λ = 200–4000 nm (Fig. S1) (Tropf and Thomas, 1997). Measuring absorption in this region is particularly challenging, as extinction is dominated by scattering.

In the absence of direct, high-sensitivity measurements of shortwave *k*, previous studies estimating the climate impacts of alumina in the stratosphere have applied varying shortwave absorption treatments to alumina particles. For example, some climate modeling studies neglect shortwave absorption by alumina (Ross & Sheaffer, 2014;

Maloney et al., 2025), while others assign values of mineral dust for simplicity (Barker et al., 2025; Jain and Hastings, 2024). Studies focused on SAI used values for shortwave *k* of ~$10^{-7}$ (Dykema et al., 2016; Keith et al., 2016; Stefanetti et al., 2024; Vattioni et al., 2025; Vattioni, Käslin, et al., 2024; Vattioni, Weber, et al., 2024). These very small values are estimated by interpolating the near-UV and infrared absorption values measured by previous studies (Tropf and Thomas, 1998) using theoretical models (Dykema et al., 2016). Importantly, these prior studies acknowledge that shortwave absorption is not accurately estimated, leading to uncertainties in their radiative impact (Dykema et al., 2016; Vattioni et al., 2025; Vattioni, Käslin, et al., 2024).

Here, we address this knowledge gap by reporting the first direct measurements of alumina aerosol *k* across the shortwave spectra using sensitive techniques (Chakrabarty et al., 2023). Absorption and scattering cross-sections were measured using five custom single-wavelength (λ = 375–1047 nm) photoacoustic spectrometers (Arnott et al., 1999, Upadhyay et al., 2025), applying first-principles methods to quantify *in situ* aerosol-phase absorption. These measurements were combined with numerically exact Mie theory–based inversions (Sumlin et al., 2018) to retrieve *k* values accurate to five decimal places. We validate the accuracy of the *in situ* measurements using electron energy-loss spectroscopy (EELS) on individual alumina nanoparticles in an aberration-corrected scanning transmission electron microscope (STEM), providing imaginary refractive index *k* values from λ = 50 to 1200 nm. Finally, we estimate the radiative impacts of alumina using the SOCOL-AERv2 (Vattioni, Weber, et al., 2024) aerosol chemistry climate model to understand the potential consequences of the finite shortwave absorption measured in this study.

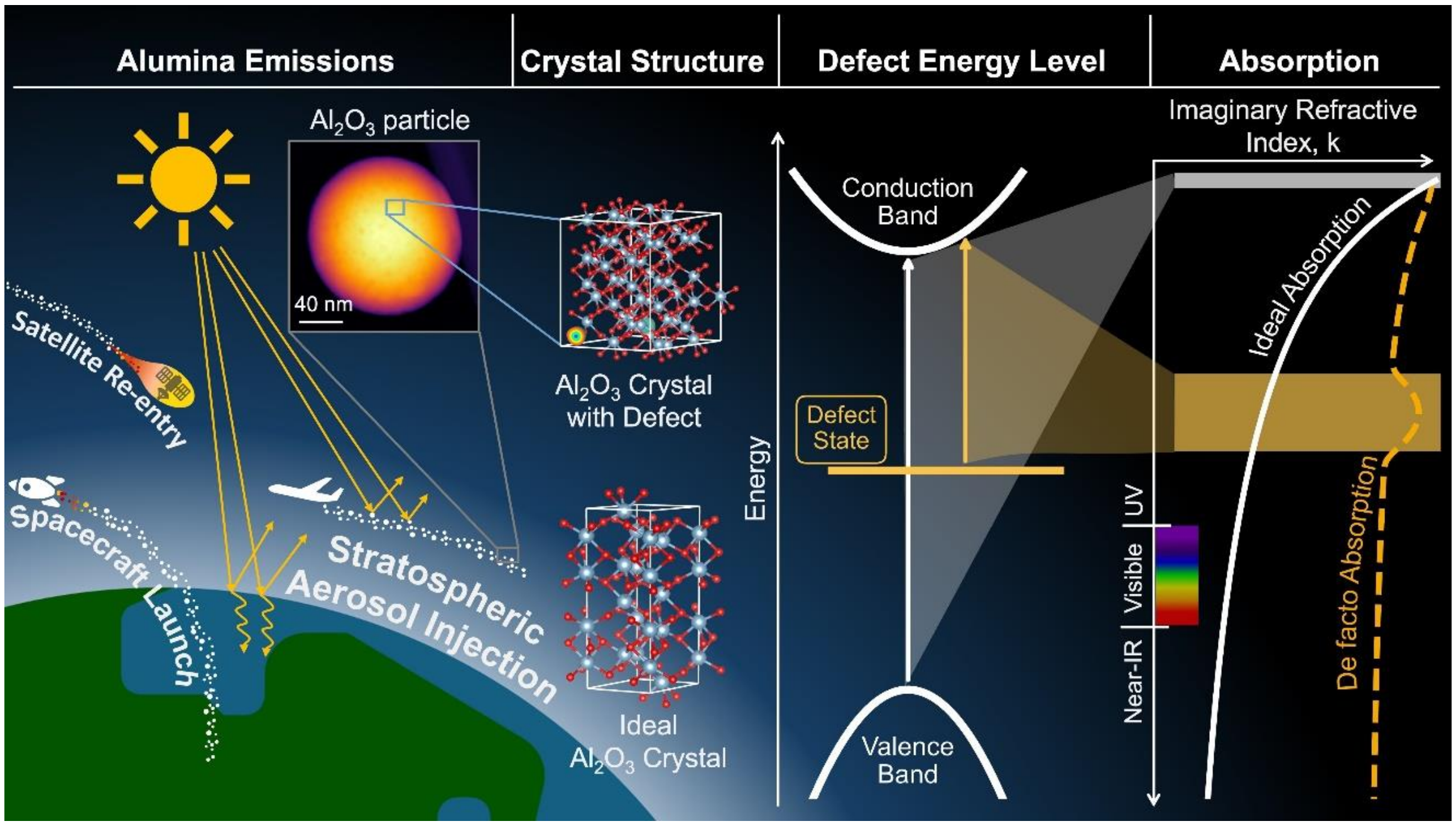

Figure 1: **Sources and optical behavior of alumina aerosol.** Alumina is emitted into the stratosphere from spacecraft launches and is independently proposed as a SAI-based geoengineering particle. While hypothetical pristine crystalline alumina may exhibit negligible absorption below its bandgap energy (8.7 eV), entropy-driven defects (for example, oxygen vacancies) can introduce sub-bandgap states, leading to non-negligible sub-bandgap absorption.

## 2. Methods

### 2.2 Experimental Setup

Three types of polydisperse alumina nano-powder were tested: spherical-shaped particles of $\alpha$-phase, irregular-shaped particles of $\alpha$-phase, and irregular-shaped particles of $\gamma$-phase (US Research Nanomaterials Inc., USA) (Table S1). Each powder was individually aerosolized using a dry dispersion method adapted from a previous study (Mitroo et al., 2021). The bench-scale dust generator (Fig. 2) consisted of a custom-built Erlenmeyer flask, containing the alumina powder, which was agitated using a wrist-action shaker (Model 75, Burrell Scientific LLC, USA) and periodically tapped with a metal rod to minimize particle agglomeration. Zero air (UN1002, AirGas, Inc., USA) was introduced tangentially at the bottom of the flask at a variable flow rate of 1-6 L min$^{-1}$, as the carrier gas. The resulting aerosol was diluted with air and passed through a $PM_1$ cutoff cyclone (SCC1.829, MesaLabs, Inc., USA) to remove coarse particles. The aerosol stream was then neutralized by $^{210}$Po strips (housed in the Brechtel 9000) to eliminate electrostatic charges generated during dispersion and sent to a buffer chamber to ensure uniform aerosol mixing before simultaneous sampling by downstream instruments.

Aerosols drawn from the buffer chamber were used to measure the real-time particle number size distributions using a Scanning Mobility Particle Sizer (SMPS 3938, DMA 3082 and CPC 3787, TSI Inc., USA) and an Aerodynamic Aerosol Classifier (AAC, Cambustion Ltd, UK). *In situ* absorption ($\beta_{abs}$) and scattering ($\beta_{sca}$) coefficients were measured at five wavelengths (λ = 375, 405, 532, 721, and 1047 nm) using bespoke Integrated Photoacoustic Nephelometers (IPNs), described in the following paragraph.

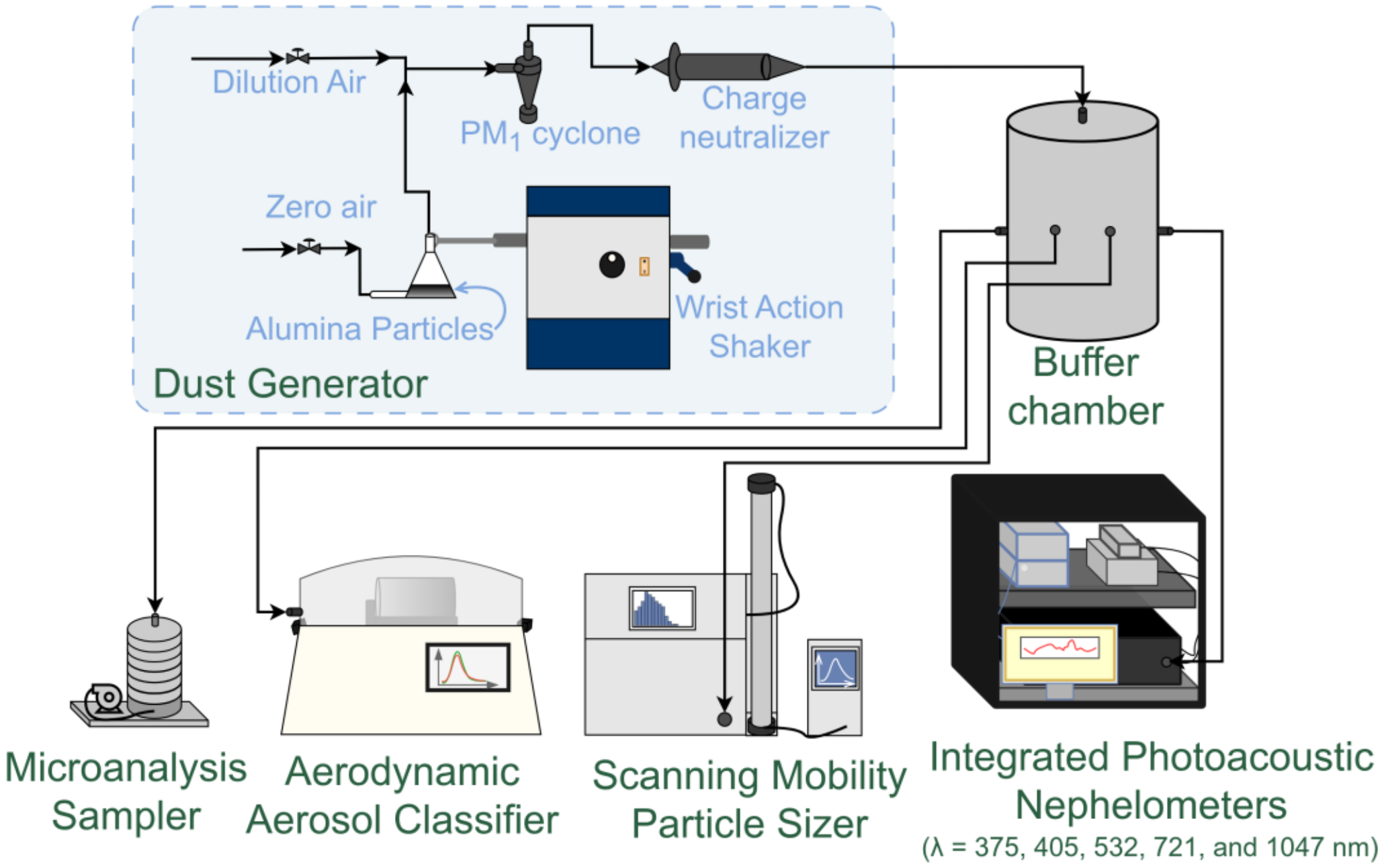

**Figure 2: Experimental setup to measure the shortwave absorption properties of alumina particles.** It consists of a constant output dust generator (adapted from Mitroo

et al., 2021), which supplies aerosols to a buffer chamber from which aerosols are sampled by *in situ* aerosol measurement instruments and deposited on electron microscopy grids using a microanalysis sampler.

### 2.3 Photoacoustic spectrometry and Mie-theory based inversions

The IPNs sampled aerosols at a flow rate of 1 L $min^{-1}$ to simultaneously measure wavelength-dependent absorption $\beta_{abs}(\lambda)$ and scattering $\beta_{sca}(\lambda)$ coefficients via photoacoustic spectrometry (Arnott et al., 1999; Upadhyay et al., 2025) and integrating nephelometry (Varma et al, 2003), respectively. Upon entering the instrument, aerosol particles are illuminated by an amplitude-modulated laser beam where they simultaneously scatter and absorb light. The scattered light is detected by a photodiode, providing the scattering coefficient. The absorbed energy manifests as heat in the surrounding gas, inducing pressure waves (or sound). The intensity of these pressure waves, which is proportional to the modulated laser power and aerosol absorption, is detected in real-time by a sensitive microphone and used as a direct measure of aerosol absorption coefficient ($\beta_{abs}$).

The complex refractive index ($m = n + ik$) of the particles was retrieved by inverting the measured $\beta_{abs}(\lambda)$, $\beta_{sca}(\lambda)$ using the simultaneously measured particle number size distributions (Fig. S2). For a given size distribution and a predefined range of real ($n$) and imaginary ($k$) refractive index values, we generated two-dimensional heatmaps of scattering and absorption coefficients using forward Mie-theory calculations (using the PyMieScatt python package) (Sumlin et al., 2018). A search algorithm then identified contour lines on the heatmaps corresponding to the experimentally measured $\beta_{abs}(\lambda)$ and $\beta_{sca}(\lambda)$ values (Sumlin et al., 2018). The intersection of these two contours yielded the unique solution for the complex refractive index ($n$, $k$). This directly inverted value was used for spherical $\alpha$-alumina particles. For non-spherical particles (irregular $\alpha$ and $\gamma$), the Maxwell Garnett mixing rule (Heinson & Chakrabarty, 2016) was employed to estimate the bulk refractive index, treating irregular particles as inclusions embedded in an air medium. The real part of refractive index for the Maxwell Garnett calculations was the same as that reported by Tropf and Thomas (1997) (1.75-1.79), similar to the measured *n* for spherical particles (1.69-1.75).

For each particle type, multiple measurements of the refractive index were made at all wavelengths. These included 3–6 measurements for spherical α-phase particles, 25–38 measurements for irregular α-phase particles, and 39–45 measurements for irregular *γ*-phase particles at each of the five wavelengths. Measurements at 721 nm for *γ*-phase were excluded due to a laser malfunction in the photoacoustic spectrometer. For each particle type, the median value of the refractive index at a given wavelength was calculated from all measurements. The mean and standard deviation of these median values across the three particle types are shown in Fig. 3 and were used to fit the data to an Urbach tail model (Moosmüller et al., 2011).

### 2.3 Electron Energy Loss Spectroscopy

Samples for transmission electron microscopy were prepared by drawing aerosols from the buffer chamber and depositing onto lacey carbon-coated copper grids (#1881, Ted Pella Inc., USA) using a microanalysis sampler (MPS-6, California Measurements Inc., USA). Prior to characterization in the electron microscope, the TEM grids were baked under vacuum at 160 °C for 8 hours to remove surface contamination. We estimate these conditions correspond to a temperature of approximately 465 °C at atmospheric pressure, using the Clausius–Clapeyron equation (Chakrabarty et al., 2023)— that is sufficiently below the melting point of alumina (Tropf and Thomas, 1997).

The EELS measurements were performed using the Nion UltraSTEM 100 operated at Oak Ridge National Laboratory, equipped with a Nion Iris spectrometer. The microscope was operated at an accelerating voltage of 100 kV employing a probe convergence semi-angle of 30 mrad. EEL spectra were acquired using a collection semi-angle of 35 mrad and a dispersion of 50 meV per channel over 1028 channels, corresponding to an energy window of approximately -4 to 36 eV.

EEL spectra were acquired from the center of each nanoparticle, irrespective of its lateral size. The local thickness used in the Kramers-Kronig analysis was estimated from the particle diameter based on geometric considerations. All spectra were calibrated by aligning the zero-loss peak (ZLP) to 0 eV using the center-of-mass algorithm as implemented in the Nion Swift package. The reflected tail method was employed to extract the ZLP, with a measured full width at half maximum (FWHM) of approximately 0.3 eV. Fourier-log deconvolution was subsequently applied to extract the single scattering distribution (SSD) using Digital Micrograph. Accurate SSDs could be extracted down to ~1 eV (corresponding to a wavelength of ~1240 nm). A cosine-bell function was used to smoothly interpolate below 1 eV. The complex dielectric function was then derived from the SSD using Kramers-Kronig analysis (Chakrabarty et al., 2023).

### 2.4 Aerosol Chemistry Climate Model Simulations

The geophysical impacts of the measured shortwave absorption were estimated using the aerosol chemistry climate model SOCOL-AERv2 (Stenke et al., 2013, Feinberg et al., 2019). The model is based on the ECHAM5.4 general circulation model (Roeckner et al., 2006), the MEZON chemistry module (Rozanov et al., 1999; Egorova et al., 2003), and the AER aerosol microphysics module (Weisenstein et al., 1997, 2007). The model has a horizontal resolution of 2.8° x 2.8° (T42), a vertical resolution of 39 sigma pressure levels from the surface up to 0.1 hPa altitude (about 80 km) and dynamical, chemical and microphysical timesteps of 15 minute, 2 hours and 3 minutes respectively. The model employs time sliced yearly repeating conditions of 2020 concentrations of greenhouse gases (SSP5-8.5, O'Neill et al., 2015) and ozone depleting substances (WMO, 2018) as well as a climatological 10-year (2010–2019) averaged seasonal cycle of sea surface temperatures (SSTs) and sea ice concentrations (SICs), from the Hadley dataset (Kennedy et al., 2019).

SOCOL-AERv2 has recently been extended to represent solid particle aerosol microphysics (Vattioni, Weber et al. (2024) and has been used extensively to simulate SAI scenarios of solid aerosol particles (Stefanetti et al., 2024; Vattioni, Käslin, et al., 2024; Vattioni et al., 2025). Our simulations follow the same configuration as that of Vattioni, Käslin et al., (2024), but with the updated shortwave imaginary refractive index reported in the present study. We simulated uniform injection of 5 Mt year$^{-1}$ of alumina particles between 30° N and 30° S at 54 hPa altitude (about 20 km) and the results presented are averages of three ensemble members with 15-year steady state conditions (45 years of model data). Monodisperse particles were injected with a radius of 215 nm, close to the optimal size for SAI applications (Dykema et al., 2016; Vattioni, Käslin et al., 2024). The optical property tables input to the model were updated for both monomers (Bohren and Huffman, 2008; Mätzler, 2002) and aggregates (Rannou et al., 1999), to accurately represent the optical properties of alumina particles. To isolate the impact of shortwave absorption, two sets of simulations were performed: one including the measured shortwave absorption and one in which shortwave absorption was neglected.

## 3. Results

### 3.1 Shortwave Imaginary Refractive Index

*In situ* aerosol photoacoustic spectrometry and integrating nephelometry confirmed that the alumina particles were predominantly scattering, with single scatter albedos ranging 0.997-0.999 across the shortwave spectrum. The aerosolized particles had a lognormal number size distribution with median diameters between 200 and 500 nm and geometric standard deviations between 1.5 and 1.7 (Table S1). The imaginary indices ranged from $1.4 \times 10^{-4}$ to $1.2 \times 10^{-3}$ across shortwave wavelengths (375–1047 nm; Fig. 3) for the three types of particles, including α- and γ-phase powders. The measured *k* did not differ substantially between α- and γ-phase particles (Fig. S3), and averaged values for three aerosol types are shown in Fig. 3 (green scatter markers).

The shortwave imaginary *k* for *α*-alumina particles was independently validated using particle-scale EELS measurements. EELS measurements showed excellent agreement with the *in situ* photoacoustic spectrometry results, while extending spectral coverage to shorter wavelengths (<200 nm, orange shaded region in Fig. 3). The EELS-measured *k* exhibited an expected sharp decline below the bandgap, which is also observed in previous measurements (Tropf & Thomas, 1997), and is consistent with the bandgap theory of insulators. However, instead of a sharp decline in *k,* there is a prominent absorption ridge between 207-310 nm (4-6 eV), centered at 240 nm (5.18 eV), shown in Fig. S4. This absorption ridge is likely associated with a mid-bandgap energy state, likely attributable to oxygen vacancy defects in the α-alumina crystal structure (Pustovarov et al., 2010). These defect-induced states likely contribute to the substantial absorption observed in this study with *k* values over three orders of magnitude higher than previous theoretical estimates for idealized alumina crystals (Dykema et al., 2016; Tropf & Thomas, 1997). Moreover, this unexpectedly large absorption by alumina nanoparticles uncovers the need to investigate the absorption properties of other candidate aerosol materials proposed for stratospheric aerosol injection (Dykema et al., 2016; Keith et al., 2016; Pope et al., 2012; Vattioni, Käslin et al., 2024; Vattioni et al., 2025; Weisenstein et al., 2015).

The *in situ k* measurements were fitted with an Urbach energy model, commonly used to estimate absorption by crystalline materials (Moosmüller et al., 2011). The resulting fit and associated uncertainty (green shaded area in Fig. 3) align well with the bandgap edge and show good agreement with the shortwave EELS data. This fit enables a continuous *k* spectrum from 200 to 6000 nm, which, when combined with Tropf and Thomas' (1997) ultraviolet and infrared data (Fig. 3), facilitates modeling of alumina aerosol climate impacts.

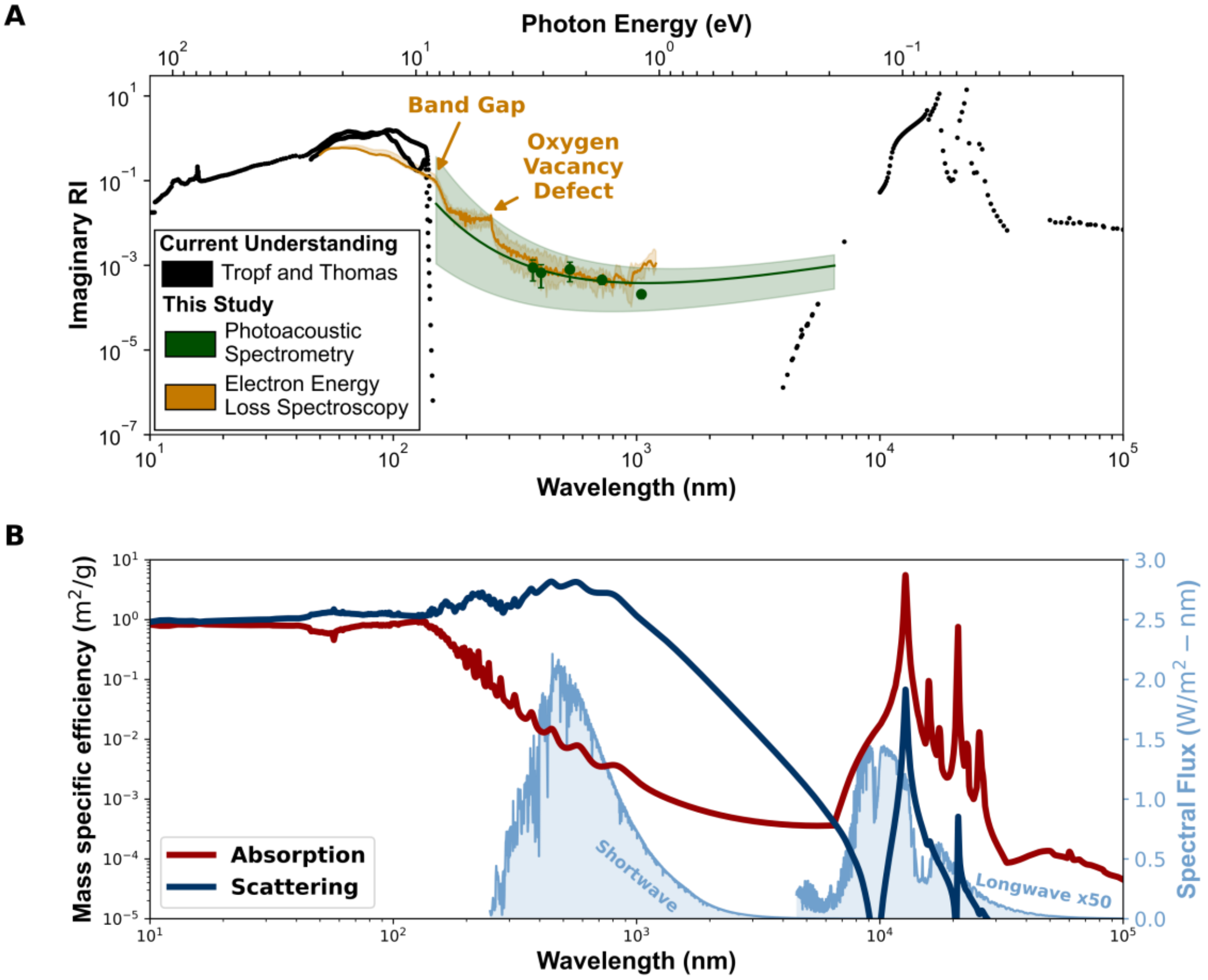

**Figure 3:** **(A)** Spectral variation of the imaginary refractive index (*k*). Tropf and Thomas (1997) reported *k* across a wide spectral range (10–$10^5$ nm), but values between 200 and 4000 nm (shortwave solar spectrum) are missing. We report shortwave *k* using photoacoustic spectrometry (green markers, mean ± standard deviations of three particle types) and electron energy loss spectroscopy (orange; shaded region represents 25$^{th}$ to 75$^{th}$ percentile of ten particles). The *k* values measured using photoacoustic spectrometry were fit with an Urbach tail model (Moosmüller et al., 2011), $k = a\lambda \exp(b/\lambda)$, yielding $a = (1.28 \pm 0.58) \times 10^{-7}$ and $b = 1094 \pm 196$ nm (uncertainties are the 25$^{th}$ to 75$^{th}$ percentile values, green shaded area). **(B)** Spectral variation in the mass absorption and scattering efficiencies. Scattering dominates extinction at shorter wavelengths, and absorption dominates extinction at longer wavelengths. The measured finite shortwave absorption

is orders of magnitude lower than the shortwave scattering and longwave absorption. The blue colored shaded region in the background shows the top-of-the-atmosphere incoming shortwave solar radiation and outgoing infrared longwave radiation (right y-axis).

### 3.2 Impact of Shortwave Absorption

Simple calculations of shortwave absorption radiative forcing using the method developed by Ross & Sheaffer (2014) (Supplementary Section S1), lead to an atmospheric forcing ranging from 5-225 mW $m^{-2}$ (linearly increasing with atmospheric concentrations) (Fig. S5). This forcing exceeds the forcing by black carbon particles from rocket launch emissions (4.4 mW $m^{-2}$) (Ryan et al., 2022), currently the largest anthropogenic forcing agent in the stratosphere. However, these calculations of instantaneous absorption may not present a holistic picture, since atmospheric adjustments could mask any increased absorption. Further, alumina exhibits substantial absorption in the atmospheric window (Tropf & Thomas, 1997), which overshadows the shortwave absorption observed in this study (Fig. 3b).

To assess alumina's potential radiative and climatic impacts, we ran aerosol chemistry climate model (SOCOL_AERv2) simulations with the updated alumina absorption properties and compared them with simulations with no shortwave absorption. The introduction of the defect-induced shortwave absorption did not yield a substantial change in the globally averaged net top-of-atmosphere (TOA) radiative forcing (LW+SW), which measured -0.651±0.184 $Wm^{-2}$ with shortwave absorption versus -0.573±0.161 $Wm^{-2}$ without. Similarly, the steady state atmospheric burden remained nearly identical (4065 Gg with shortwave vs. 4058 Gg without). The vertical profile of temperature in the tropical stratosphere (Fig. 4a) shows that the observed temperature increases are primarily driven by the longwave absorption of alumina, as seen in both the tropical averages (Fig. 4a) and zonal averages (Fig. 4b). The further addition of the measured shortwave absorption does not significantly influence these temperature (Fig. 4c) or ozone concentration (Fig. 4e) fields.

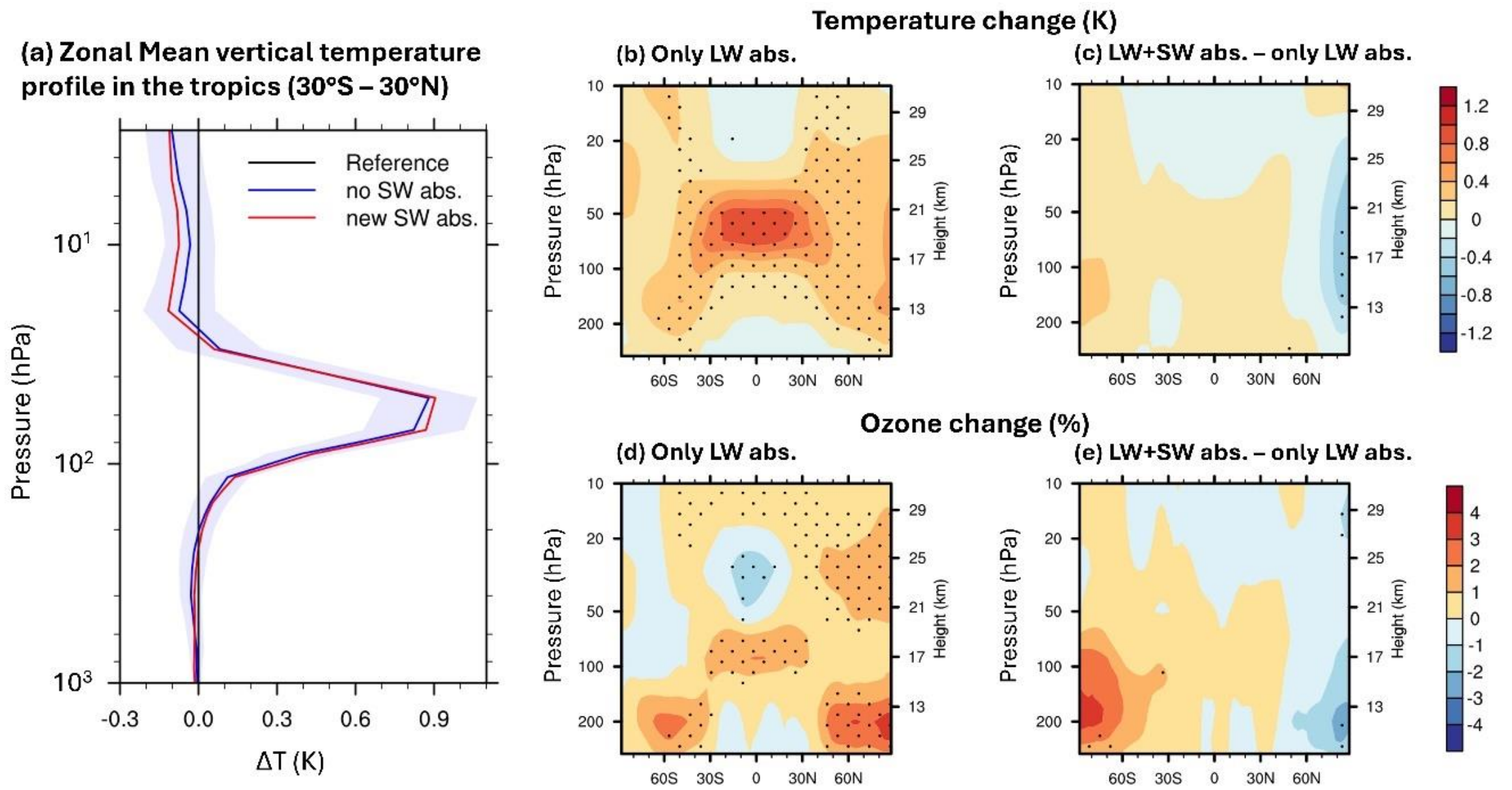

**Figure 4:** Changes in atmospheric conditions from injection of alumina particles in the stratosphere. (**a**) vertical profile of zonally averaged temperature anomaly in the tropical atmosphere (30° S to 30° N). The shaded area represents the standard deviation of inter-annual variability of the only longwave absorption scenario. Zonally averaged changes in **(b and c)** temperature (in K) and in **(d and e)** ozone concentrations (% change) of the (b and d) only longwave (LW) absorption scenario compared to the reference scenario and (c and e) the longwave and shortwave (SW) absorption scenario compared to the only longwave scenario.

## 4. **Discussion**

The presence of defects in alumina's crystal structure leads to non-trivial absorption in the shortwave spectrum. Despite the measured shortwave *k* being at least three orders of magnitude higher than previous theoretical estimates, our simulations indicate that the shortwave scattering and longwave absorption properties of alumina dominate the overall radiative effects (Fig. 3B), masking any effect of the shortwave absorption. These findings imply that alumina's shortwave absorption – a property not directly measured so far and highlighted as an uncertainty in climate model simulations – is not a major source of uncertainty in estimating the effects of alumina particles as a candidate material for SAI. However, substantial uncertainties remain concerning other properties which will need to be addressed. For example, large uncertainties persist in the heterogeneous reactive uptake of stratospheric chlorine reservoir species on alumina surface under stratospheric conditions, that could influence ozone depletion (Vattioni et al., 2023, Vattioni et al., 2025; Huynh and McNeill, 2024; Weisenstein et al., 2015). Furthermore, background climatic conditions can also influence stratospheric transport, and consequently the radiative impacts of SAI deployment (Vattioni et al., 2025).

While globally averaged effects on the radiative balance do not seem to change appreciably with the introduction of shortwave absorption in SAI scenarios of 5 Mt/yr of alumina particles, subtle changes in zonal mean water vapor and temperature are observed (Fig. S6). These localized changes, particularly those at the sensitive north and south poles, could be accentuated in other SAI scenarios, such as the potentially more practical high-latitude low-altitude scenarios (Robock et al., 2008; Lee et al., 2021). Further, scenarios involving seasonal injections at select locations (Visioni et al., 2019) could influence meridional stratospheric temperature gradients sufficiently to cause detectable downstream hydrological feedback (Visioni et al., 2019; Wang et al., 2018; Ferraro et al., 2015). Similarly, the calculations of alumina's radiative effects from increasing emissions due to rocket launches and spacecraft reentries (Barker et al., 2024; Brown et al., 2024; GAO, 2022, Revell et al., 2025), will also require revisiting. Emissions from these sources are typically confined in the mid-latitudes, primarily in the northern hemisphere (Barker et al., 2024; Maloney et al., 2025; Ryan et al., 2022).

The likely insignificance of alumina's enhanced shortwave absorption is a result of the dominance of two other factors, both dictated by the particle's refractive indices: relatively strong longwave absorption and strong shortwave scattering. Alumina exhibits relatively strong absorption in the atmospheric window (8-14 µm) of outgoing longwave absorption, with a significant absorption peak in $k$ at ~13 µm (Fig. 3). Concurrently, its high shortwave real refractive index (~1.77) results in strong shortwave scattering. While a magnitude of $k$~$10^{-4}$ is insignificant for alumina, it may be substantial for other candidate materials. Unlike alumina, which exhibits a significant disparity between its shortwave scattering and longwave absorption, other species may have weaker absorption within the atmospheric window. In such cases, an absorption index of $k$~$10^{-4}$ could become significant, necessitating independent examination of every candidate SAI particle. Currently, the shortwave $k$ of other weakly absorbing SAI candidate materials also does not have measured shortwave indices owing to measurement difficulties or are based on measurements of ideal thin-film crystals (Dykema et al., 2016) that may not be representative of practically produced nanoparticles. Their shortwave $k$ may also be influenced by crystal defects or impurities potentially causing deviations from presumed idealized behavior. Therefore, the optical properties of other candidate materials, which exhibit a variety of imaginary and real refractive indices, need to be determined on a case-by-case basis to understand whether shortwave absorption is significant. The sensitive measurement techniques employed in the present study could facilitate the detection of weak, yet potentially significant, shortwave absorption signals.

## 5. Conclusion

Solid particles are proposed for SAI deployment due to their advantages over sulfate aerosols, notably their weaker longwave absorption and more efficient shortwave scattering. Alumina is one such candidate with desired properties, including its presumed weak absorption in the shortwave spectrum. We present the first direct measurements of this hitherto elusive shortwave absorption using sensitive photoacoustic and EELS techniques, finding the imaginary indices to range from $1.4\times10^{-4}$ to $1.2\times10^{-3}$ across the shortwave spectrum. This absorption is over three orders of magnitude higher than the values assumed by previous climate model studies that evaluate their SAI radiative impacts. The finite shortwave absorption observed in this study is likely attributable to oxygen vacancy defects in alumina's crystal lattice, which lead to sub-bandgap absorption with a peak at 240 nm. Such defects are thermodynamically favorable and thus unavoidable during nanoparticle formation. This work confirms that imperfections in crystal structures can indeed lead to unexpected, non-trivial absorption, a phenomenon previously theorized as a major source of uncertainty.

We assessed the impact of the measured shortwave absorption by using the measured *k* values as input into the SOCOL-AERv2 aerosol chemistry climate model. The simulations revealed that the addition of the shortwave absorption does not appreciably alter the radiative balance in simulations of SAI scenarios and thus, neither stratospheric temperature nor ozone concentrations. This indicates that alumina's shortwave scattering and longwave absorption are far stronger processes, dominating the overall radiative budget and masking any effect of the enhanced shortwave absorption. This finding implies that the uncertainty surrounding alumina's shortwave absorption is not a major limiting factor in the assessment of its potential use as an SAI material. However, despite reducing this major uncertainty, other uncertainties such as heterogeneous chemistry on alumina particles (Vattioni et al., 2023), ecological impacts (Vattioni et al., 2024) or technical feasibility (Smith et al., 2020) are still large and need to be constrained (Vattioni et a. 2023, Vattioni et al., 2025). Furthermore, the measured *k* values may not necessarily be insignificant for other SAI candidate materials, owing to differences in their intrinsic absorption and scattering properties. Hence, systematic measurement of other candidate species remains prudent to accurately constrain their radiative impacts.

**Data Availability**

Imaginary refractive index data, processing codes, and intermediate EELS analyses are available at Kapoor et al. (2025). The SOCOL-AER aerosol chemistry climate model code is available via Vattioni (2023). The aerosol chemistry climate model outputs will be made available on an online repository upon acceptance of the manuscript.

**Acknowledgments**

We acknowledge the Simons Foundation (grant nos. SFI-MPS-SRM-00005174, SFI-MPS-SRM-00005217 and SFI-MPS-SRM-00005208) for project funding. We thank Dr Huafang Li for helping acquire electron microscopy images at Washington University and the Institute of Material Science and Engineering, Washington University in St Louis for the use of the Scanning Transmission Electron Microscope. Scanning transmission electron microscopy at Oak Ridge National Lab was supported by the Center for Nanophase Materials Sciences (CNMS), which is a US Department of Energy, Office of Science User Facility at Oak Ridge National Laboratory.

**Author Contributions**

Conceptualization: RKC, RM, LX. Methodology: TSK, PU, DM, RKC, RM. Investigation: TSK, PU, JH, GR, JAH, SV, JD, JS, RKC, RM. Visualization: TSK, PU, JH, GR, SV, RKC, RM. Supervision: RM, RKC. Writing—original draft: TSK, PU, RKC. Writing—review editing: TSK, PU, JH, GR, JC, DM, SV, JD, JS, JK, JAH, LX, RKC, RM.

**Supporting Information for**

# Unveiling the Shortwave Absorption Spectra of Alumina Aerosols: Implications for Solar Radiation Modification

Taveen Singh Kapoor[a,1], Prabhav Upadhyay[a,1], Jian Huang[b,1], Guodong Ren[b], John Cavin[b], Dhruv, Mitroo[c], Sandro Vattioni[d,e], John Dykema[f], Jan Sedlacek[e], Joshin Kumar[a], Jordan A. Hachtel[g], Lu Xu[a], Rohan Mishra[b,*], Rajan K. Chakrabarty[a,*]

[a] Center for Aerosol Science and Engineering, Department of Energy, Environmental and Chemical Engineering, Washington University in St. Louis, St. Louis, USA.

[b] Institute of Materials Science & Engineering, Department of Mechanical Engineering and Materials Science, Washington University in St. Louis, St. Louis, USA.

[c] Veterans Affairs Medical Center, St. Louis, USA.

[d] Institute for Atmospheric and Climate Science, ETH Zurich, Zurich, Switzerland

[e] Physikalisch-Meteorologisches Observatorium Davos and World Radiation Center, Davos, Switzerland.

[f] John A Paulson School of Engineering and Applied Sciences, Harvard University, Cambridge, MA, USA

[g] Center for Nanophase Materials Sciences, Oak Ridge National Laboratory, Oak Ridge, USA.

[1] These authors contributed equally

* Correspondence to Rajan K. Chakrabarty or Rohan Mishra

**Email:** chakrabarty@wustl.edu; rmishra@wustl.edu

This file contains:

Figures 1 to 6

Table 1

## S1: Method for simple forcing calculations:

### Instantaneous Shortwave Absorption Radiative Forcing calculations

The Instantaneous shortwave absorption radiative forcing ($RF_{ISWA}$) was calculated using the method developed by Ross and Sheaffer (2014) Briefly, emitted alumina particles are assumed to accumulate in the stratosphere between latitudes 30° N and 80° N (with accumulation area, *A* in $m^2$) for four years. Therefore, the steady state atmospheric burden (*M*, Gg) is taken as four times the emissions. Shortwave absorption radiative forcing is calculated using equation 1, where $I_{SW}(\lambda)$ is the global and time averaged spectral solar shortwave flux (W $m^{-2}$ $nm^{-1}$) and $MAC(\lambda)$ is the mass absorption cross-section ($m^2$ $g^{-1}$).

$$RF_{ISWA} = \frac{M}{A}\int_{250\,nm}^{3000\,nm} I_{SW}(\lambda)\,MAC(\lambda)d\lambda \qquad (1)$$

MAC is calculated using the Urbach tail fit (Moosmüller et al., 2011) of the photoacoustic spectroscopy measured imaginary refractive indices measured in the present study and the particle size distributions employed to estimate heterogenous ozone loss on alumina, reported by Danilin et al. (2001) An average of cases B and D in Danilin et al. (2001) were taken as plausible alumina size distribution for rocket launches, with the assumption that only *Dp*<1$\mu$m remain suspended longer in the stratosphere. For stratospheric aerosol injection scenarios, a log-normal size distribution with a median diameter of 430 nm (considered ideal for SAI; Dykema et al., 2016) and geometric standard deviation of 1.5 was taken. The calculated MAC using the two particle number size distributions were nearly identical and hence those reported by Danilin et al. (2011) were used throughout. MAC was calculated using equation 2.

$$MAC(\lambda) = \frac{\int_0^{1\mu m} C_{abs}(k,\lambda,D_p)n(D_p)dD_p}{\int_0^{1\mu m} \rho\pi D_p^3 n(D_p)dD_p/6} \qquad (2)$$

where $C_{abs}$ ($m^2$) is absorption cross-section (calculated using Mie theory), $n(D_p)$ ($cm^{-3}$ $nm^{-1}$) is the number size distribution, $D_p$ (nm) is the diameter and $\rho$ (g $m^{-3}$) is the density of the aerosol. The spectral (200-6000 nm) MAC value may be obtained using the following fit:

$$\log_{10}(MAC) = A_0 + A_1\log_{10}\left(\lambda/_{550}\right) + A_2\left[\log_{10}\left(\lambda/_{550}\right)\right]^2 + A_3\left[\log_{10}\left(\lambda/_{550}\right)\right]^3 \qquad (3)$$

Where $\lambda$ is wavelength in nm and $A_0$ = -0.88 (-0.78 – -1.01), $A_1$ = 2.07 (1.73 – 2.34), $A_2$=-2.13 (-1.97 – -2.64), $A_3$ = -2.3 (-2.72 – -2), where the ranges in brackets correspond to uncertainties (25$^{th}$ to 75$^{th}$ percentile) associated with uncertainties in the imaginary refractive index.

## Figures

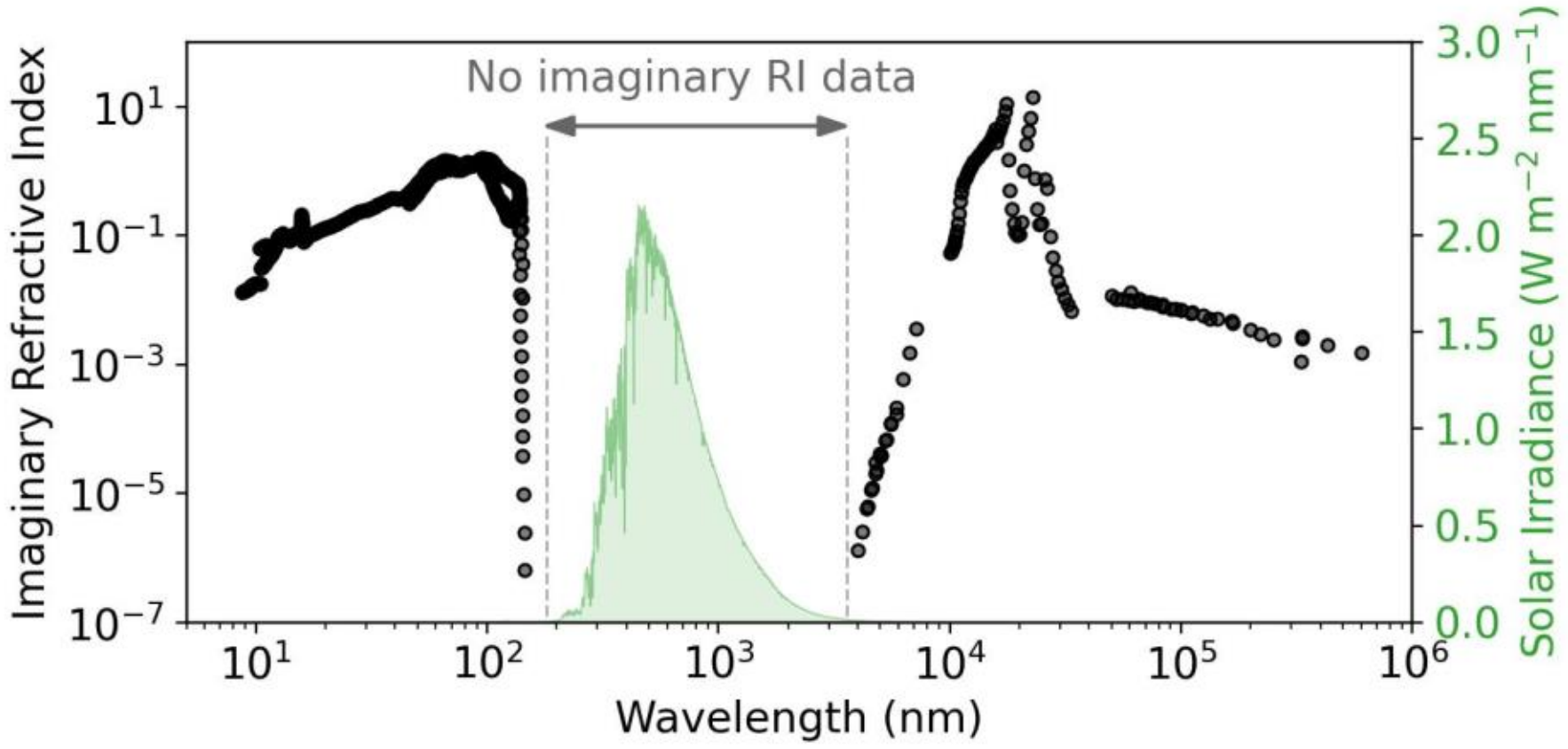

Figure S1: Imaginary refractive index reported by Tropf and Thomas (1997) (left y-axis) and the top-of-the-atmosphere incoming shortwave solar radiation (right y-axis). No measurements are reported in the shortwave spectrum.

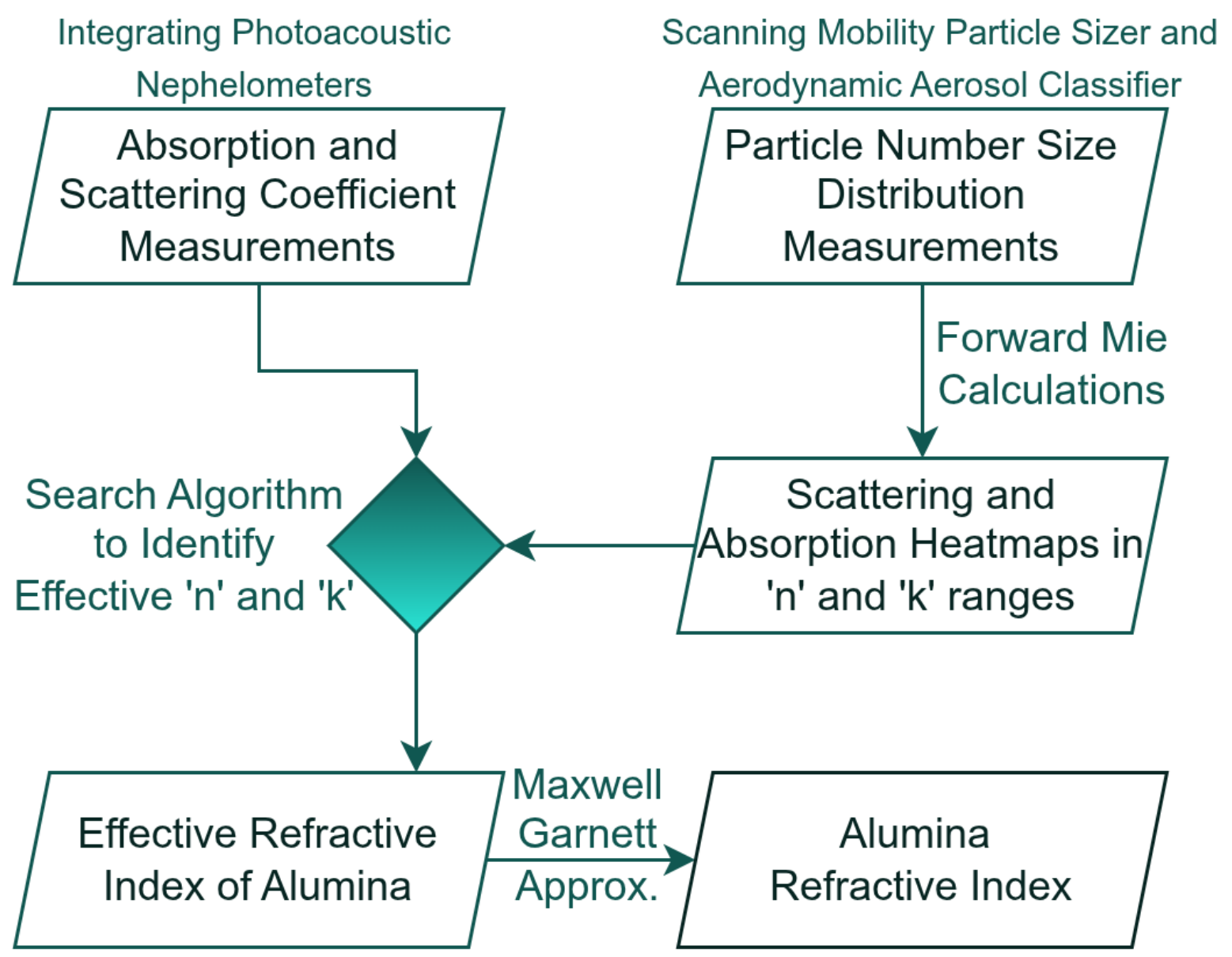

Figure S2: Refractive index inversion technique flowchart.

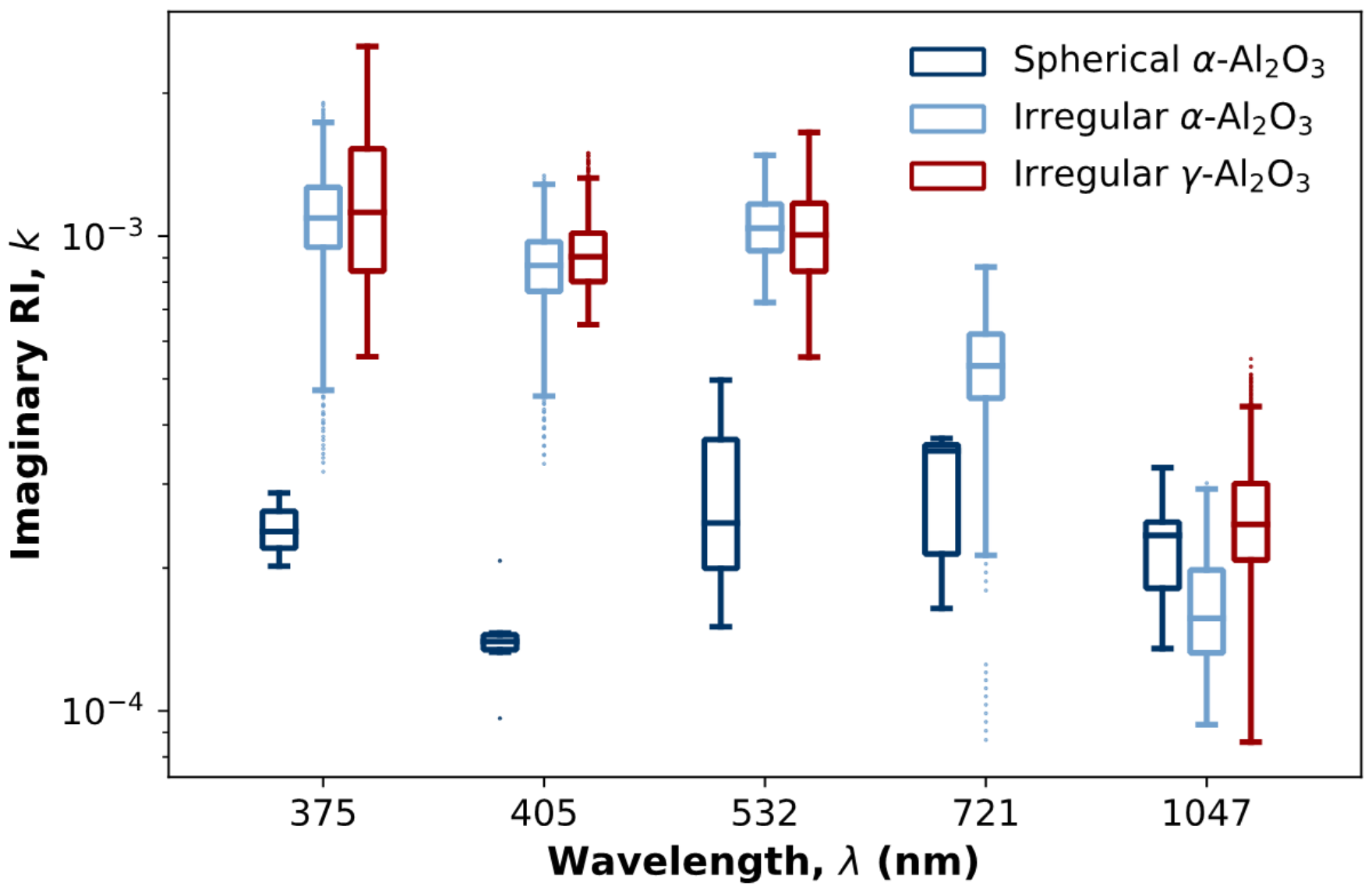

Figure S3: Imaginary refractive index retrieved from photoacoustic spectrometry and Mie-inversions.

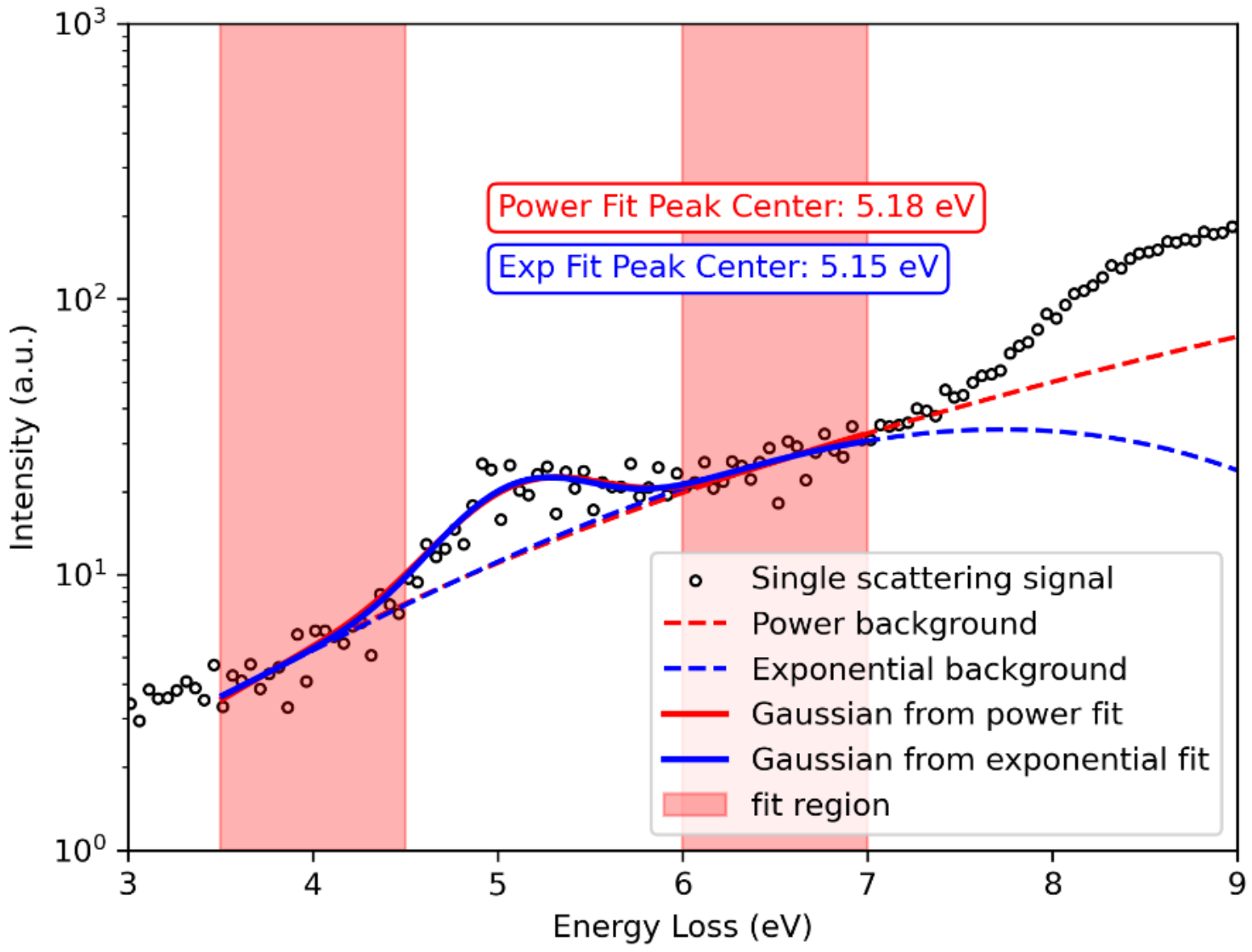

Figure S4: Peaks fitted to the single scattering signal from EELS. The background has been fit with a power law (red) and a third-order exponential (blue). The two fits are almost identical and overlap. The extracted defect peaks using the two different backgrounds are shown around 5.18 eV.

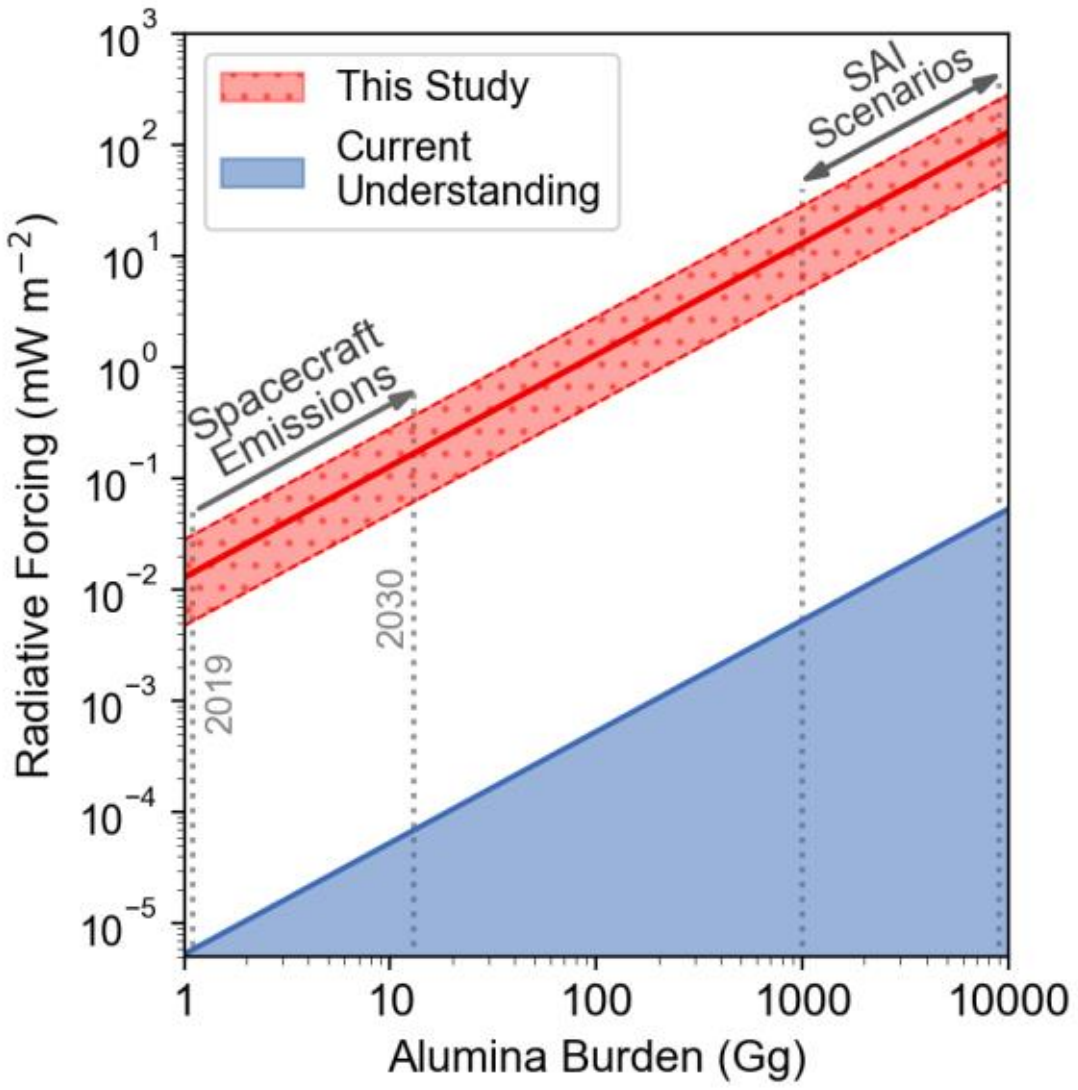

Figure S5: Shortwave absorption radiative forcing of alumina for present and projected and stratospheric aerosol injection (SAI) burdens. The solid red line represents forcing calculated using

measured refractive indices (*k*); dashed lines indicate uncertainties associated with measured *k*. The blue shaded region shows forcings computed using previously assumed *k* bounds ($k_{550}$ = 0–$10^{-7}$).

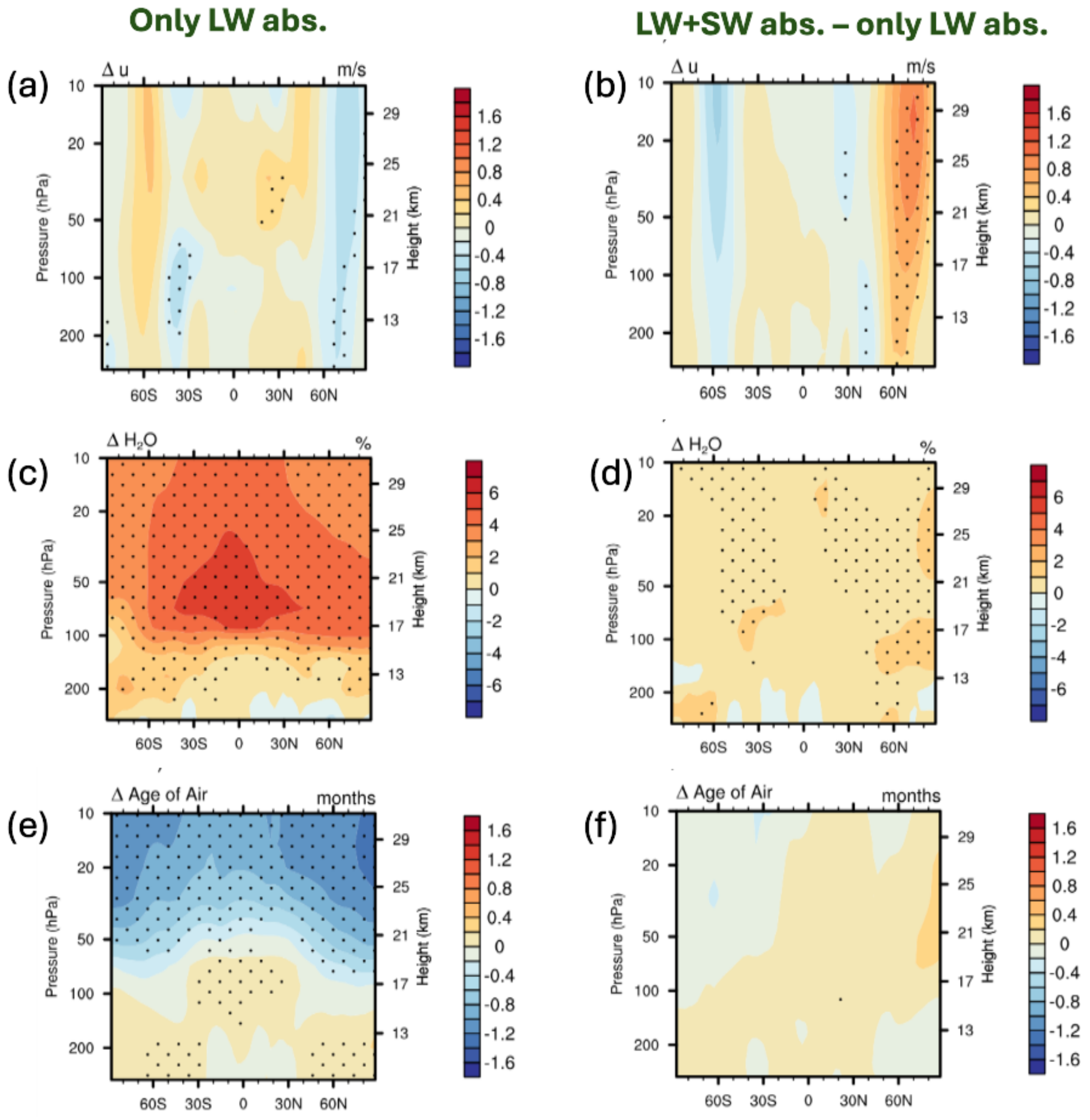

Figure S6: Incremental changes in zonal mean wind speeds (Δu), water vapor pressure ($\Delta H_2O$), and age of air (Δ Age of Air) with introduction of longwave and shortwave absorption of alumina as simulated by the SOCOL_AeRv2 aerosol chemistry climate model.

## Tables

Table S1: Details about alumina nano powders used and aerosols generated.

| | α-spherical | α-irregular | γ-irregular |
|---|---|---|---|
| Manufacturer Stock # | US1160M | US1129M | US1171M |
| Phase | Alpha | Alpha | Gamma |
| Shape | Spherical | Irregular | Irregular |
| Mean Diameter (nm) | 310-510 | 240-350 | 175-420 |
| Geometric Standard deviation | 1.46-1.92 | 1.51-1.67 | 1.22-1.81 |